# The Secret Life of a Process Description: A Look Into the Evolution of a Large Process Model

Martín Soto
Alexis Ocampo
Jürgen Münch

Fraunhofer Institute for Experimental Software Engineering
Fraunhofer-Platz 1
67663 Kaiserslautern, Germany
{soto, ocampo, muench}@iese.fraunhofer.de

**Abstract:** Software process models must change continuously in order to remain consistent over time with the reality they represent, as well as relevant to the task they are intended for. Performing these changes in a sound and disciplined fashion requires software process model evolution to be understood and controlled. The current situation can be characterized by a lack of understanding of software process model evolution and, in consequence, by a lack of systematic support for evolving software process models in organizations. This paper presents an analysis of the evolution of a large software process standard, namely, the process standard for the German Federal Government (V-Modell® XT). The analysis was performed with the Evolyzer tool suite, and is based on the complete history of over 600 versions that have been created during the development and maintenance of the standard. The analysis reveals similarities and differences between process evolution and empirical findings in the area of software system evolution. These findings provide hints on how to better manage process model evolution in the future.

**Keywords:** process modeling, process model change, process model evolution, model comparison, V-Modell® XT

## 1 Introduction

In his seminal paper from 1987 [1], Leon Osterweil pointed out the similarities between software processes and software programs. 20 years later, however, it is clear to us that his vision of process descriptions similar in its degree of formality and detail to actual computer programs has been much harder to realize than he actually envisioned. In fact, the majority of contemporary, practical software process descriptions still contain a large proportion of informal material in the form of natural language text. This does not mean, however, that process descriptions must be completely informal. Indeed, they are often highly structured and use standardized, uniform terminology. They also often contain an intricate lattice of internal and external cross-references that are not only intended to guide the reader in navigating the description but also ensure the description's internal consistency. The presence of this complex internal structure, and the consistency requirements associated with it, clearly make process descriptions look similar to software systems in many respects.

One aspect of this analogy that has undergone little research until now is the evolution of large process descriptions and its relation to the much better understood field of software evolution. As every process modeling practitioner can attest to, changing a process description over time while preventing its structure from deteriorating or its consistency from being lost is a difficult task. Still, it remains unclear up to what extent maintaining a software process description is similar to maintaining a software system, and how much of the existing software maintenance knowledge can be extrapolated to the process realm. While considering this fundamental question, a number of more concrete questions may arise, for instance:

- if an evolving process description increases its complexity over time unless work is done to reduce it;
- if most changes of process models are performed only on a few parts of a process description;
- if changes performed shortly before a release cause more post-release changes than changes performed earlier before a release;
- if parts of process models that have been changed many times have a higher probability of additional changes.

We expect the answers to such questions to be useful for supporting process management activities better than they can be supported nowadays. Knowing, for example, that changing certain areas of a process description may potentially imply further changes in the near future, could be used to inspect these changes more carefully or to avoid changing certain parts of a process description for minor reasons.

Our current knowledge of process model evolution is not sufficient to answer these questions on an empirical basis. This is caused, in part, by the fact that mechanisms and tools for analyzing process model evolution and visualizing the results are widely missing. Another reason is that only few organizations have a history of the versions of their process models in sufficient detail, that is, including versions in between releases and documented justifications (i.e., rationale) for the changes introduced in each new version.

In this article, we present preliminary results aimed at understanding process model evolution. Our findings are based on detailed evolution data for a large and complex process description: the German V-Modell® XT. This description is interesting not only because of its significance for the German information technology domain, but also because of its large size and complexity. The V-Modell describes about 1500 process entities, and its printed documentation is over 700 pages long.

In order to perform our analysis, we applied novel comparison and annotation techniques to identify the changes made to the model over its versioning history, and to link these changes, whenever possible, with their underlying rationale. By doing this, we obtained a comprehensive, integrated representation of the V-Modell's life along three major public releases and over 600 individual versions. With this information as a basis, we have been able to answer a number of basic questions related to the V-Modell's evolution. These questions, as well as the way we approached them, form the core of this article.

The rest of the paper is structured as follows: Section 2 gives a short overview of the evolution of the V-Modell® XT. Section 3 briefly discusses the techniques used to perform our analysis of the model. Section 4 presents our analysis in more detail, and discusses its results.. The paper closes with an overview of related work, a summary, and an outlook on future work.

## 2  The German V-Modell® XT and the History of its Evolution

The German process standard V-Modell [2] (not to be confused with Royce's V-Model [3]) has a long history, and an ever increasing significance for the German IT landscape. Its origin dates to the mid-eighties. In 1997, the so-called V-Modell 97 was officially released as a software development standard for the German federal government. The standard remained unchanged until 2004, when a consortium of industrial and research institutions received public funding to perform a thorough update of the model. The result was the new V-Modell® XT, which was established as German federal standard for software development. Since its inception, the model has seen continuous updates, represented by three major and two minor releases. Also, since a few months ago, an English version has also been available, which is kept synchronized with the original German version.

The V-Modell® XT is a high-level process description, covering such aspects of software development as project management, configuration management, software system development, and change management, among others. In printed form, the latest English version at the time of this writing (version 1.2.1) is 765 pages long and describes about 1500 different process entities.

Internally, the V-Modell® XT is structured as a hierarchy of process entities interconnected by a complex graph of relationships. This structure is completely formalized, and suitable for automated processing. The actual text of the model "hangs" from the formalized structure, mainly in the form of entity and relationship descriptions, although a number of documentation items (including a tutorial introduction to the model) are also integrated into the structure in the form of *text module* entities. Actual editing of the model is performed with a software tool set created specially for the purpose. The printed form of the V-Modell® XT is generated automatically by traversing the structure in a predefined order and extracting the text from the entities found along the way.

The V-Modell® XT contents are maintained by a multidisciplinary team of experts, who work, often concurrently, on various parts of the model. In order to provide some measure of support to this collaborative work, the model is stored as a single XML file in a standard code versioning system (CVS). As changes are made by the team members, new versions are created in this system. As usual for a versioning system, versions can, and often do, include a short comment from the author describing the changes. Also, an Internet-based issue tracking system is available so that model users can report problems with the model. This system often includes discussions between team members and users about how certain issues should be resolved. Not all actual changes in the model can be traced to a particular issue in the tracking system, but many of them can.

The change logs show that, since its initial inception, the model has been changed often and for a wide variety of reasons. Changes can be as simple as individual spelling or grammar corrections, or as complex as the introduction of a whole set of processes for hardware development and software/hardware integration. The richness and complexity of this change history makes the V-Modell a very interesting target for evolution analysis.

## 3   Analyzing the Evolution of a Process Description

The first step in order to analyze the evolution of this process description was to read its versioning history into our Evolyzer model comparison system. Although a description of the internal operation of Evolyzer is beyond the scope of this paper (see [4] for details), a short explanation of its workings is in order. The basis of the system is a model database that can contain an arbitrary number of versions of a model. The formalism used for representing the models is the RDF notation [5] and the whole model database can be queried using a subset of the SPARQL [6] query language for RDF.

The central characteristic that distinguishes Evolyzer from other RDF storage systems is its ability to efficiently compare versions of an RDF model. Given two arbitrary versions, the system is able to compute a so-called *comparison model* that contains all model elements (RDF statements, actually) present in the compared versions, marked with labels indicating whether they are common to both versions, or are only present in one of them, and, in the latter case, in which one of the versions they are present. Given the high level of granularity of this comparison, identifying changes in it by direct inspection is generally a difficult task. For this reason, change identification is performed by looking for special *change patterns* in the comparison model (see [4] for a detailed explanation.) This not only makes it possible to look for changes that are specific, in their form or structure, to a particular model schema, but allows for restricting change identification to particular areas of the model or to specific types of model elements.

For the present study, we attempted to read 604 versions from the original versioning repository into our system. These versions were created in somewhat more than two years time, with three major and one minor public releases happening during that period. Since Evolyzer uses the RDF notation for model representation (this is necessary in order for our comparison technique to work at all), each V-Modell version was mechanically converted from its original XML representation into an RDF model before reading it into the system. This conversion did not add or remove information, nor did it change the level of formalization of the original process description. Process entities described in the original XML through XML elements were translated into RDF resources (the original XML already contained unique identifiers, which were reused for the RDF resources) and the text associated to them was stored as RDF property values. Relations encoded in XML as element references were converted into RDF relations. The conversion process was successful for all but 4 of the 604 analyzed versions. These 4 versions could not be read into our repository because their corresponding XML files contained syntax errors.

After importing the version history, we proceeded to compare the versions pairwise to identify individual changes happening from one version to the next. As changes, we considered the addition or deletion of entities, the addition or deletion of relations between entities, and the alteration of text properties. We identified these changes by defining corresponding change patterns and searching for them in the version comparisons. Information about each of the identified changes including type, version number and affected process entities was encoded in RDF and stored in the repository together with the model versions. This allowed us to easily go from the change information to the actual model contents and back from the models to the changes as necessary for our analysis (see [7] for the details of how this cross referencing works.)

## 4   An Exploratory Look Into a Process Description's Evolution

The resulting RDF repository containing detailed information about the V-Modell's change history provided the basis for our exploratory analysis of the model's evolution. Our long-term research goal is to formulate explicit verifiable hypotheses about process model evolution, but in order to do that, observation is a first, indispensable step. For this reason, the fundamental objective of the present analysis was to observe and informally characterize the evolution of the model. We attempted to do that by formulating rather open questions and then trying to extract data from the change repository and visualize them in such a way that we could attempt to address the questions by direct observation.

Given the complex structure of the V-Modell® XT, we concentrated our analysis on only one part of it, namely, the so-called *process modules,*[1] a number of large entities that act as containers for a good number (but not all) of the finer-grained entities in the model. We did this for two reasons. First, the process modules contain the "meat" of the description, namely, the process entities used to describe the actual process steps and products: activities, products, roles, and the relationships among them. Second, since process modules are the official means for tailoring the model to specific project types, they correspond to sensible components of the whole description, and are thus more likely to produce meaningful results when observed independently from each other.

Additionally, and for the sake of simplicity, we decided to reduce this analysis to changes affecting the text descriptions contained in the entities, and to exclude the relationships connecting entities. In the following, we present the analysis questions, together with the approach we took to analyze them, the resulting visualization, and the results we derived from it.

---

[1] In German, process modules are called *Vorgehensbausteine,* a term that would rather correspond to *process building blocks*. We decided, however, to stick to the translation used by the official English version of the V-Modell® XT.

### 4.1 Complexity Over Time

The starting point of the analysis is the question of whether the V-Modell has increased its complexity over time. This question is related to Lehman's law with respect to system evolution, which states that the complexity of a system increases over time unless work is done to reduce it ([8], cited in [9]). To address this question, we chose a simple metric for the model complexity, namely, the total number of entities contained in each process module. By running a special query and performing simple postprocessing of the results, we determined this number for each process module and for each of the 604 analyzed versions, and produced individual plots displaying the process module's size for each version number. Due to space limitations, we are omitting the individual plots (22 in total) but Figure 1 shows the total size accumulated over the 22 process modules for each version number.

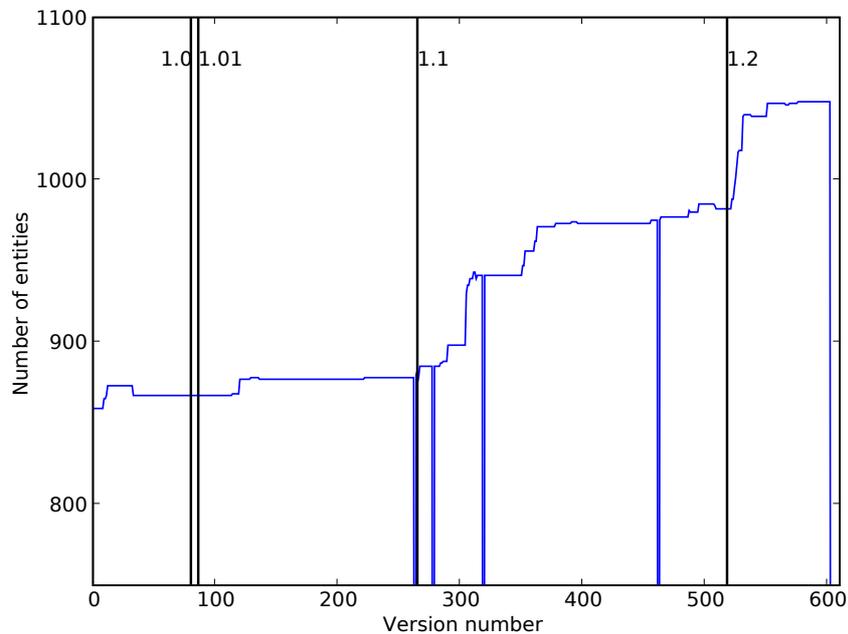

**Fig. 1.** Number of entities in the process modules along the version history.

The curve in Figure 1 shows a clear growing tendency, going from around 850 to over 1000. Pronounced growth is observed after two of the releases, probably pointing to major changes that were held back until the release. The analysis of the plots covering specific process modules (not included here) shows a similar growing tendency. Significant reductions of the entity count can only be observed in cases where a module was split at some point. As the cumulative graph shows, however, this did not affect the total element count. Some "dents" can be observed at the 4 points were versions could not be read.

Even despite major restructuring, the total number of entities in the V-Modell® XT increased significantly during the observed period. This growth can be attributed, at least in part, to model enhancements such as the introduction of processes for hard-ware development. Still, these results suggest that monitoring the complexity of process descriptions and possibly taking measures to keep it under control can be a valuable strategy for maintaining complex process models.

### 4.2 Distribution of Changes Over Time and Over the Model

The next questions are concerned with the way changes affect different parts of the model: How are changes distributed among versions, and how do they relate to releases? How are they distributed over the process modules?

Our approach to addressing these questions was to display the changes in such a way that the process module affected by the change, as well as the time and size of the changes, become visible. Figures 2 and 3 are two such displays.

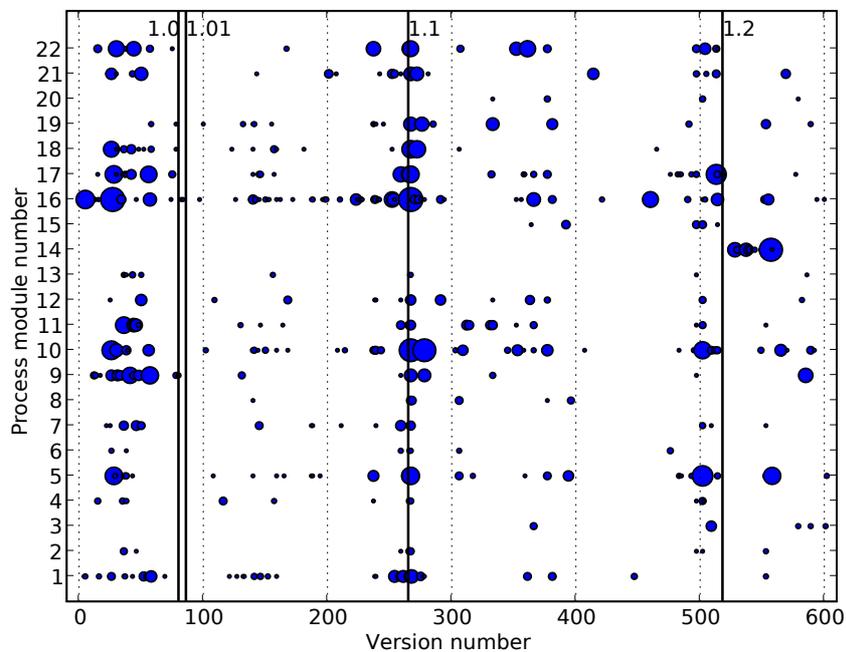

**Fig. 2.** Changes discriminated by process module along the version history

The X-axis in Figure 2 shows the version number (from 1 to 604), whereas the Y-axis shows the process modules (numbered arbitrarily). There is a circle wherever a version changed an entity in a particular process module. The size of the circle is proportional to the number of changes affecting entities in the module. Figure 3 is similar to Figure 2, but the X-axis corresponds to actual calendar times. Changes are displayed at the locations where their corresponding versions were checked in. Since versions are not distributed uniformly across time, this figure also contains a "version density" bar at the bottom that has black bars at the points in time where versions actually happened.

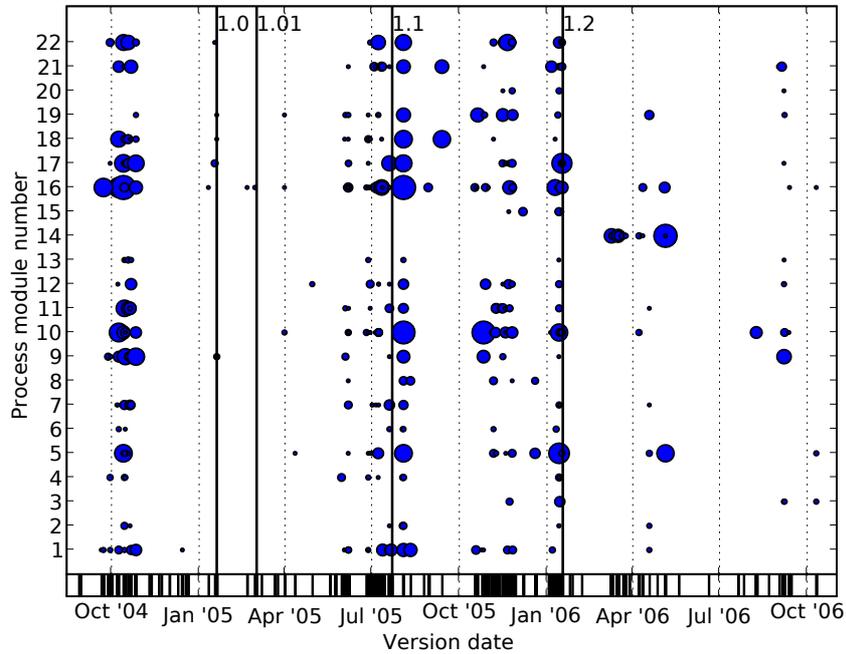

**Fig. 3.** Changes discriminated by process module, against time.

Several points are worth mentioning about these figures. First, activity concentrates around releases. Activity before a release probably corresponds to release preparation, whereas activity after a release points to changes held back and introduced after the release. Both of these points were corroborated verbally by members of the V-Modell development team, and can also be confirmed by inspecting the version logs. An interesting observation is that the version-based graph (Figure 2) also looks busy around releases, implying that versions close to a release often collect more changes in a single version than versions far from the release. If this were not the case, the "congestion" around releases would only be observable on the time-based graph. A partial explanation for this phenomenon is that a number of the versions grouping several changes are related to reviews or to other bulk corrections that result in many small changes spread over the model. Still, it is possible that some of the "congested" versions are the result of changes being rushed into the model shortly before a release.

One aspect that is evident on the time-based graph is that release 1.0 looks delayed with respect to its preparatory "burst" of activity, with a period of low activity before the actual release date. According to the team members, the bulk of the preparatory work was done for the model's "presentation in society" at a public event in November 2004. Only minor corrections were made until the official release in February 2005. This can also be observed by looking at the version density bar, which shows clear peaks of activity around the releases, except for release 1.0.

Finally, the version-based graph shows us a number of process modules that present more activity than the remaining ones: 10, 16, 5 and, to some extent, 1. Although this is often related to their larger size, change activity seems not to be strictly proportional to size, and also seems to depend on the relative importance of the various modules (we need to investigate both of these points in more detail). The graphs also show that process modules often undergo "bursts" of activity that calm down later on, such as the one observed in process module 10 between releases 1.1 and 1.2. This suggests that, similar to what happens in software development, complex changes have to be performed step-wise and often introduce errors that must be corrected later.

The previous observations point in different ways to the similarities between process model and software evolution. In particular, one should not believe that change management is simpler or less risky for process models than it is for software systems. Practices such as inspections, configuration management, or issue management are most probably advisable for complex modeling projects and may even be necessary in order to achieve high-quality results over time.

### 4.3 Changes in Detail

Our last question is concerned with the relationship between local and global changes: Does the evolution of individual modules look similar to the evolution of the whole model? To address this question, we decided to analyze the change history of one single process module in more detail.

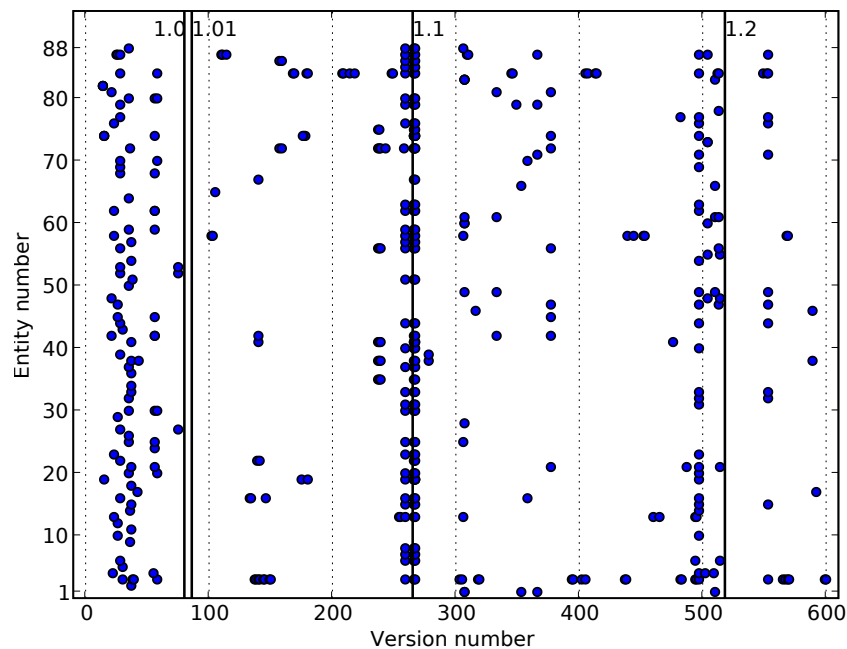

**Fig. 4.** Changes in process module *System Development.*

Figure 4 shows the changes happening to single entities in one particular module, *System Development*, (number 10 in the previous two figures). The X-axis corresponds to the version number, whereas the Y-Axis shows the entity number (entities were numbered arbitrarily). A dot is present where a version changes a particular entity.

The first observation is that this figure presents a pattern similar to the one in Figure 2, with changes concentrating around the releases. Also, shortly before the releases, there are versions affecting many entities at a time. This corroborates a similar observation made in the previous section.

An interesting point is that several instances of changes happening in sequence to a particular entity can be seen on the graph. Although the change logs show that this may happen for a variety of reasons, it would be interesting to analyze whether it has a statistical significance, that is, when an entity was changed, there is a higher probability that it will be changed in the near future. Also, it would be interesting to determine which types of changes are more likely to cause such follow-up changes. Knowing this would make it possible to better handle similar changes in the future.

## 5 Related Work

Several other research efforts are concerned, in one way or another, with comparing model variants syntactically, and providing an adequate representation for the resulting differences. Most of them, however, concentrate on UML models representing diverse aspects of software systems. Coral [10], SiDiff [11], UMLDiff [12] and the approach discussed in [13] deal with the comparison of UML models. Although their basic comparison algorithms are applicable to our work, they are not concerned with providing analysis or visualization for specific uses. Additionally, FUJABA [14] manages model versions by logging the changes made to a model during editing, but is not able to compare arbitrary model versions. Models must also be edited with the FUJABA tool in order to obtain any useful change information.

Mens [15] presents an extensive survey of approaches for software merging, many of which involve comparison of program versions. The surveyed works mainly concentrate on automatically merging program variants without introducing inconsistencies, but not, as in our case, on identifying differences for analysis. The Delta Ontology [16] provides a set of basic formal definitions related to the comparison of RDF graphs. SemVersion [17] and the approach discussed by [18] are two systems currently under development that allow for efficiently storing a potentially large number of versions of an RDF model by using a compact representation of the raw changes between them. These works concentrate on space-efficient storage and transmission of change sets, but do not go into depth regarding how to use them to support higher-level tasks (such as process improvement).

We are not aware of any previous work on analyzing the evolution of process descriptions.

## 6 Summary and Future Work

Software process descriptions are intended to be faithful representations of the actual processes used to develop and maintain software systems. This fact implies a twofold challenge for process engineers: On the one hand, descriptions must be continuously improved in order to make them closer to the actual process and to make them more accessible to their users. On the other hand, as processes are improved and expanded to deal with new development challenges, descriptions must be changed accordingly. We have used novel tools and techniques to gain some insight into the evolution of a large, practical process description. We expect to use the results of the initial observations, such as those presented here, for formulating specific hypotheses to guide our future research.

A number of research directions seem promising. Currently, we are in the process of analyzing the results of the connection of the V-Modell's change history with two sources of information related to the rationale of the changes: the human edited version log and the issue tracking system [7]. We expect this to give us more insight into the dynamics of the change process: what causes changes in the first place and how the various motivations for change affect process descriptions at different points in their evolution. In particular, this may help us identify areas of the process that may be continuously causing problems, so that future improvement efforts can concentrate on them.

A final, more general question is related to process adoption. The introduction of good practices to an organization's software process involves complex learning and an increase in the necessary level of discipline. For this reason, finding an appropriate strategy for introducing good practices over time in a non-disruptive, coherent way can be very difficult. We consider that studying process evolution may teach us how to effectively introduce good practices into new organizations or into groups within an organization.

**Acknowledgments.** We would like to thank Rubby Casallas from Los Andes University, Bogotá, Colombia, for her valuable comments on a draft of this article. We would also like to thank Sonnhild Namingha from Fraunhofer IESE for proofreading this paper. This work was supported in part by the German Federal Ministry of Education and Research (V-Bench Project, No. 01| SE 11 A).